\documentclass[conference]{IEEEtran}
\IEEEoverridecommandlockouts
\usepackage{epsfig,epstopdf,setspace,subfigure}
\usepackage{amsmath,amssymb,cite,enumerate}
\newtheorem{proposition}{Proposition}

\begin{document}
\title{On the Gaussian Z-Interference Channel with Processing Energy Cost}
\author{\IEEEauthorblockN{Xi Liu and Elza Erkip\\} 
\IEEEauthorblockA{ECE Department\\
 Polytechnic Institute of NYU\\
 Brooklyn, NY 11201, USA}
\thanks{This material is based upon work partially supported by NSF Grant No.
0635177, and by CATT of Polytechnic Institute of NYU.}}
\maketitle

\begin{abstract}
This work considers a Gaussian interference channel with processing energy cost, which explicitly takes into account the energy expended for processing when each transmitter is on. With processing overhead, bursty transmission at each transmitter generally becomes more advantageous. Assuming on-off states do not carry information, for a two-user Z-interference channel, the new regime of very strong interference is identified and shown to be enlarged compared with the conventional one. With the interfered receiver listening when its own transmitter is silent, for a wide range of cross-link power gains, one can either achieve or get close to the interference-free upper bound on sum rate.
\end{abstract}
\section{Introduction}
In wireless communications, many devices such as mobile phones and hand-held computers are often limited by batteries. However, there are many demands on the limited battery, such as CPU processing, networking activity and screen backlight\cite{processingcost:Vallina10}. It is often the case that a substantial fraction of the total energy expended by the transmitter of a battery-limited device other than being radiated in transmission is for processing the signals in circuits \cite{processingcost:Wang_Chandrakasan02}\cite{processingcost:Balasubramanian09}. This makes the study on the implications of power consumption beyond transmit power meaningful. In particular, for nodes operating in the power-limited regime, it is important to take into account ``processing energy'', the energy spent for processing, in capacity calculations.

The impact of processing energy for communicating over an additive white Gaussian noise channel was first studied by Youssef-Massaad et al. in \cite{processingcost:Massaad_Medard_Zheng04}, where the circuit of a transmitter is modeled as consuming some fixed power when it is in the ``on'' state. It was shown there that when the processing energy per channel use is large, bursty Gaussian signaling achieves capacity. The authors also extended their analysis to an M-user multiple access channel (MAC) \cite{processingcost:Massaad_Medard_Zheng_Allerton}\cite{processingcost:Massaad_Medard_Zheng_TWC08} and showed that time division multiple access outperforms other schemes in terms of the sum rate.

This paper studies the impact of processing energy costs at the transmitter for a two-user interference channel (IC). The two-user IC is a fundamental model in information theory for studying interference in communication systems, in which two senders transmit independent messages to their corresponding receivers via a common channel. When processing overhead is not negligible, bursty transmission which is optimal in the point-to-point communication has the potential to reduce the effect of interference. For instance, we may schedule user bursts to overlap as little as possible to avoid interference or let a receiver hear the interferer's signals when its own transmitter is silent for the purpose of interference decoding. This paper explores how to make best use of the burstiness of both users' transmission to minimize the rate loss due to processing overhead over the IC.

We focus on the simplest IC model, the Gaussian Z-IC (also known as the one-sided IC), in which one receiver receives an interference-free signal while the other receives a combination of the intended and the interfering signals. For the case of no processing energy cost, the Gaussian Z-IC has been extensively studied in the literature, and constitutes one of the few examples of an IC for which the capacity region or the sum capacity have been established for all values of channel parameters\cite{processingcost:Sato81}\cite{processingcost:Sason_IT04}. 

Focusing on strategies that do not utilize on-off states of transmitter for signaling, we show that as cross-link power gain $a$ gets sufficiently large, each user can achieve a rate as the single-user case. The exact channel conditions for this phenomenon are identified and compared with the no processing cost case. For a general Gaussian Z-IC with arbitrary $a$, we propose several simple joint transmission schemes including a non-bursty transmission scheme, time division multiplexing (TDM) and others. The sum-rate performances of these schemes are analyzed and compared with an no-interference upper bound. We show that even with simple schemes, one can achieve or get very close to the upper bound for very small or large $a$'s.

The remainder of this paper is structured as follows. In Section II, we present the system model and relevant assumptions used. For the Gaussian Z-IC with processing overhead, conditions for the new regime of very strong interference are identified in Section III. In Section VI, we investigate how to maximize the sum-rate performance through the use of several simple joint transmission schemes for a general Gaussian Z-IC with an arbitrary $a$. Finally we draw conclusions and discuss future directions in Section V.

\section{System Model}
\begin{figure}
\centering
\includegraphics[width = 2.1in]{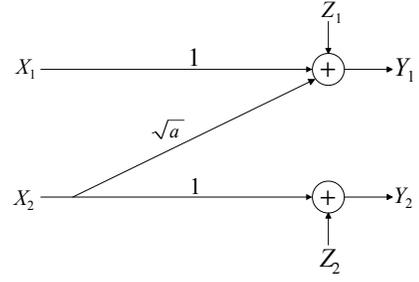}
\caption{Gaussian Z-IC}
\label{fig:fig1}
\end{figure}

 In this paper, we focus on the two-user Gaussian Z-IC as shown in Fig. \ref{fig:fig1}
\begin{eqnarray}
Y_{1,t} = X_{1,t} + \sqrt{a}X_{2,t}+Z_{1,t}\\
Y_{2,t} = X_{2,t} + Z_{2,t}
\end{eqnarray}
where $X_{i,t}$ and $Y_{i,t}$ represent the input and output of user $i \in \{1,2\}$ at time $t$, respectively, and $Z_{1,t}$ and $Z_{2,t}$ are i.i.d. Gaussian with zero mean and unit variance. Receiver $i$ is only interested in the message sent by transmitter $i$. Each user is subject to a maximum average power constraint $P_i$. Following \cite{processingcost:Massaad_Medard_Zheng04}, the processing energy cost of transmitter $i$ is modeled as a constant amount $\epsilon_i$ ($i=1,2$) whenever transmitter $i$ is on. Besides, we assume receivers are not power limited and receiver processing overhead is thus not an issue. This scenario reflects, for example, a situation in a two-cell uplink one where both transmitters are battery-limited mobile equipments while receivers are base stations attached to a stable power supply.

For simplicity, we also assume both $P_i$ and $\epsilon_i$ are normalized by the noise variance. The power constraint at transmitter $i$ is given by
\begin{equation}
\frac{1}{n}\sum_{t=1}^n[|X_{i,t}|^2+\epsilon_i \cdot 1_{\{X_{i,t}\neq 0\}}]\leq P_i
\end{equation}
 where $1_{\{\cdot\}}$ is the indicator function.

Throughout the paper, we assume on-off states of transmitters are \emph{fixed} or \emph{deterministic}, i.e., receivers are informed beforehand about when each transmitter should be in the ``on'' or ``off'' state so that they listen when there is at least one transmitter in the ``on'' state. Note that if on-off states are not fixed ahead of time, on-off signalling such as pulse position modulation can be employed to transmit additional information. However, in this case, frequent and fast transition between ``on'' and ``off'' is required and energy cost of on-off transition cannot be neglected any more, which will cause contradiction with our assumption of constant processing energy cost. In the fixed case, each transmitter can remain in the on and off states for long durations of time and very few transitions are needed. Furthermore, encoding on-off states with information may result in non-Gaussian inputs being optimal for $X_1$ and $X_2$, further complicating code design \cite{processingcost:Zhang_Guo_isit11}.

If user $j$ ($j\neq i$) is turned off and only user $i$ transmits, by \cite{processingcost:Massaad_Medard_Zheng04}, the optimal transmission scheme for user $i$ is to let transmitter $i$ and receiver $i$ be turned on for a prescribed $\theta_i^*$  fraction of the time and use Gaussian signaling in this period of time with a fixed signal power $\nu_i^*$ where the optimal parameters $\theta_i^*$ and $\nu_i^*$ take the following forms:
\begin{equation}
\theta_i^* = \displaystyle \min\left(1,\frac{P_i W(e^{-1}(\epsilon_i-1))}{(\epsilon_i-1)(W(e^{-1}(\epsilon_i-1))+1)}\right)
\label{eqn:thetastar}
\end{equation}
\begin{equation}
\nu_i^* = \frac{P_i}{\theta_i^*}-\epsilon_i
\label{eqn:nustar}
\end{equation}

Here $W(\cdot)$ is the LambertW function. 

If $\theta_1^*+\theta_2^*\leq 1$, it is obvious that users can employ time division to avoid interference and can thus obtain interference-free rates. For any value of cross-link power gain $a$, the achievable rate region is always a rectangle, specified by the two interference-free rates. In the succeeding sections, we focus on the more interesting case $\theta_1^*+\theta_2^*> 1$ when the two users need to compete for the available channel uses.

\section{When Interference Does Not Incur Any Loss}
In this section, assuming $\theta_1^*+\theta_2^*> 1$, we determine the range of power gain $a$'s such that either user's rate does not suffer any loss due to interference. For this to be possible, transmitters of both users must employ
their optimal signaling as in the single-user case. Since there is no interference at receiver 2, user 2 can get its maximum possible rate by transmitting during $\theta_2^*$ fraction of time. In order for user 1 to achieve its interference-free rate, it must be capable of decoding user 2's codewords completely or at least finding the coded symbols overlapping with its own transmission such that after cancelation it sees an interference-free channel for at least $\theta_1^*$ fraction of time. Note that we allow receiver 1 to be on when user 2 is transmitting to enable interference decoding even though transmitter 1 may be turned off.

{\proposition The two users in the Z-IC can both achieve their maximum interference-free rates if the following condition is satisfied:
\begin{equation}
1+\nu_2^*\leq (1+a\nu_2^*)^\rho (1+\frac{a\nu_2^*}{1+\nu_1^*})^{1-\rho}
\label{eqn:condition1_1}
\end{equation}
where $\rho = \frac{1-\theta_1^*}{\theta_2^*}$. Here $\theta_i^*$ and $\nu_i^*$ represent user $i$'s optimal burst and signal power level in the interference-free case respectively.
\label{thm:thm1}
}
\begin{proof}
Using optimal signaling, the two transmitters transmit during $\theta_1^*$ and $\theta_2^*$ fractions of the time respectively and both employ Gaussian codebooks. Since $\theta_1^* + \theta_2^* > 1$, to minimize the effect of interference, their signals should overlap only for $\theta_1^*+\theta_2^* - 1$ fraction of time. Besides, receiver 1 employs successive cancelation decoding. It first decodes interference by treating its own signal as noise. After decoding and removing interference, it starts decoding its own signal. The entire transmission of user 2 can be decoded by receiver 1 so that it sees an interference-free channel after removal of interference if the following condition is satisfied,
\begin{align}
\theta_2^*\log(1+\nu_2^*) &\leq (1-\theta_1^*)\log(1+a \nu_2^*)\nonumber\\
&+(\theta_1^*+\theta_2^* - 1)\log(1+\frac{a\nu_2^*}{1+\nu_1^*})
\label{eqn:condition2}
\end{align}

Here (\ref{eqn:condition2}) implies that for user 2, the amount of information flowing in the undesired direction surpasses that flowing in the desired direction. Substituting $\rho = \frac{1-\theta_1^*}{\theta_2^*}$ into (\ref{eqn:condition2}) yields (\ref{eqn:condition1_1}) in Proposition \ref{thm:thm1}.
\end{proof}

To further simply the condition in (\ref{eqn:condition1_1}) is difficult in general. In the following, we focus on the special case of low SNR's when $\epsilon_i\rightarrow 0$, $P_i\rightarrow 0$ and $P_i/ \sqrt{2\epsilon_i} = \lambda_i$ for some constant $\lambda_i>0$, $i = 1,2$. In this case, $\theta_i^*$ in (\ref{eqn:thetastar}) can be simplified as
 \begin{equation}
 \theta_i^*=\min(1,\lambda_i)>0
 \end{equation}
Hence $\rho\geq 0$. Also, by (\ref{eqn:nustar}) it follows that $\nu_i^* \rightarrow 0$. Using Taylor series approximation, we have
\begin{equation}
1+\nu_2^*\leq (1+\rho a\nu_2^*) (1+(1-\rho)\frac{a\nu_2^*}{1+\nu_1^*})\label{eqn:condition3_1}
\end{equation}

Ignoring terms containing $(\nu_2^*)^2$ on the right side of (\ref{eqn:condition3_1}), we find the following sufficient condition
\begin{equation}
a\geq \frac{1+\nu_1^*}{1+\rho\nu_1^*}
\label{eqn:condition2_1}
\end{equation}

Furthermore, $\theta_2^*\leq 1$ gives $\rho\geq 1-\theta_1^*$. Thus, the right side of (\ref{eqn:condition2_1}) can be bounded as
\begin{align}
\frac{1+\nu_1^*}{1+\rho\nu_1^*}&\leq \frac{1+\nu_1^*}{1+(1-\theta_1^*)\nu_1^*}\\
& = \frac{(1+\nu_1^*)(1+\theta_1^*\nu_1^*)}{1+\nu_1^* + \theta_1^*(1-\theta_1^*)(\nu_1^*)^2}\\
&\leq 1+\theta_1^*\nu_1^*\\
&< 1+ \theta_1^*(\nu_1^*+\epsilon_1)\\
& = 1+P_1 \label{eqn:largervstr}
\end{align}

Remember that in the no-overhead case, if the two users transmit with average powers $P_1$ and $P_2$ respectively, $a\geq 1+P_1$ corresponds to the very strong interference regime where interference-free rates can be achieved with interference decoding. Since (\ref{eqn:largervstr}) holds, the condition in (\ref{eqn:condition2_1}) suggests that a larger range of power gains $a$ ensures a very strong IC under processing energy cost.

Depending on the values of $\lambda_1$ and $\lambda_2$, we have the following cases:
 \begin{itemize}
 \item $\lambda_i<1$ ($i=1,2$): $\nu_i^* \thickapprox \sqrt{2\epsilon_i}$ and thus $\theta_i^* = \lambda_i$; i.e., both users employ bursty transmission. Under all these assumptions, $\rho =\frac{1-\lambda_1}{\lambda_2}$. Thus, we can simplify the inequality in (\ref{eqn:condition2_1}) as
\begin{equation}
a\geq \frac{P_2+\sqrt{2\epsilon_1}P_2}{P_2+\sqrt{2\epsilon_2}(\sqrt{2\epsilon_1}-P_1)}\triangleq a_0,
\end{equation}
Note that $\theta_1^*+\theta_2^* = \frac{P_1}{\sqrt{2\epsilon_1}}+\frac{P_2}{\sqrt{2\epsilon_2}}>1$, i.e., $P_2 \sqrt{2\epsilon_1}>\sqrt{2\epsilon_2}(\sqrt{2\epsilon_1}-P_1)$; hence, $a_0>1$. We have proved $a_0< 1+P_1$ previously, thus it follows immediately that $1<a_0< 1+P_1$.
\item $\lambda_1<1$, $\lambda_2\geq 1$: in this case, we have $\nu_1^* \thickapprox \sqrt{2\epsilon_1}$, $\theta_1 = \lambda_1$, $\theta_2^* = 1$ and $\nu_2^* = P_2-\epsilon_2$; i.e., user 1 employs bursty transmission while user 2 uses the conventional transmission using the entire time period. Therefore, $\rho = 1 - \lambda_1$. Thus, the inequality in (\ref{eqn:condition2_1}) has the following simplified form:
    \begin{equation}
    a \geq \frac{1+\sqrt{2\epsilon_1}}{1+\sqrt{2\epsilon_1}-P_1} \triangleq a_1,
    \end{equation}
    Note that $a_1$ does not depend on the specific values of user 2's processing cost $\epsilon_2$ and power constraint $P_2$.
\item $\lambda_1\geq 1$: in this case, $\theta_1^* = 1$ and hence $\rho = 0$. The inequality in (\ref{eqn:condition2_1}) degenerates into the one in the trivial case $a\geq 1+\nu_1^*$ with $\nu_1^*$ equal to $P_1 - \epsilon_1$.
\end{itemize}

\section{Sum Rate Maximization}
In Section III, assuming fixed on-off states, we have characterized the optimal achievable rate region for the Gaussian Z-IC in the very strong regime in the presence of processing overhead. However, it is not easy to obtain an optimal rate region for an arbitrary $a$. Especially for $a<1$, only the sum capacity is known even in the case of no processing overhead. In this section, for an arbitrary $a$, we investigate several simple transmission schemes for the purpose of maximizing the achieved sum rate.

\subsection{Scheme I: Gaussian signaling with no burstiness}
If no burstiness is allowed and both users transmit all the time, user $i$'s signal power level would be at most $P_i-\epsilon_i$ joules per channel use. Under this assumption, when $a\geq 1$, the optimal scheme is for receiver 1 to decode interference\cite{processingcost:Sato81} and the maximum sum rate is given by
\begin{equation}
R_1+R_2 =  \min(C(P_1-\epsilon_1)+ C(P_2-\epsilon_2), C(P_1-\epsilon_1+a(P_2-\epsilon_2)))
\end{equation}
where $C(x) = \frac{1}{2}\log_2(1+x)$. Otherwise if $a<1$, treating interference as noise achieves the maximum sum rate \cite{processingcost:Sason_IT04} which is given by
\begin{equation}
R_1+R_2 = C\left(\frac{P_1-\epsilon_1}{1+a(P_2-\epsilon_2)}\right) + C(P_2-\epsilon_2)
\end{equation}

\subsection{Scheme II: Time division multiplexing (TDM)}
If two users employ TDM to avoid interference, user 1 uses $\theta_1$ ($0<\theta_1<1$) fraction of the time slots while user 2 uses the left $1-\theta_1$ fraction. It is not hard to show that it suffices to restrict $\theta_1$ to the range $1-\theta_2^* \leq \theta_1\leq \theta_1^*$. The maximum achievable sum rate can be found by solving:
\begin{equation}
\max_{1-\theta_2^* \leq \theta_1\leq \theta_1^*} \theta_1 C(\frac{P_1}{\theta_1}-\epsilon_1)
+(1-\theta_1)C(\frac{P_2}{1-\theta_1}-\epsilon_2)
\end{equation}

\subsection{Scheme III: Partial overlapping time slots, constant power}
The third scheme is to let user 1 and user 2 transmit over $\theta_1$ and $\theta_2$ fractions of the time slots respectively and allow their transmission to overlap for $\theta_1+\theta_2 -1$ fraction of time. To have a nontrivial case, the parameters need to satisfy $1-\theta_2^*\leq \theta_1 \leq 1$, $1- \theta_1^*\leq \theta_2 \leq 1$ and $\theta_1+\theta_2\geq 1$. Each user still uses a constant signal power level during its entire transmission. Besides, independent information is sent by user 2 during $1-\theta_1$ fraction of the time when it transmits alone and $\theta_1+\theta_2-1$ fraction of the time when it interferes with user 1. Hence, receiver 1 does not need to be on when transmitter 1 is off. In the duration $\theta_1+\theta_2-1$, if $a>1$, using the strong interference results \cite{processingcost:Sato81}, we can achieve the best sum rate during overlap by decoding the interference. Therefore, for $a\geq 1$, the maximum sum rate can be found by solving:
\begin{align}
\max_{\begin{subarray}{c}\theta_1,\theta_2\end{subarray}}&\; (1-\theta_2)C(\nu_1) + (1-\theta_1)C(\nu_2) \nonumber\\
&+ (\theta_1+\theta_2-1)\min(C(\nu_1)+ C(\nu_2),C(\nu_1+a\nu_2))
\end{align}
where $\nu_1 = P_1/\theta_1 -\epsilon_1$ and $\nu_2 = P_2/\theta_2 -\epsilon_2$.  A similar reasoning applies for $0\leq a<1$ by treating interference as noise and the sum rate can be obtained as
\begin{equation}
\max_{\begin{subarray}{c}\theta_1,\theta_2\end{subarray}}\;(1-\theta_2)C(\nu_1) + \theta_2 C(\nu_2)+ (\theta_1+\theta_2-1)C(\frac{\nu_1}{1+a\nu_2})
\end{equation}
Scheme II is a special case of Scheme III when $\theta_1$ and $\theta_2$ are chosen such that $\theta_1 + \theta_2 = 1$ while Scheme I is another special case when $\theta_1 = \theta_2 = 1$.
\subsection{Scheme IV: Partial overlapping time slots, constant power, joint decoding over the entire time period ($a\geq 1$ only)}
Recall that in Section III, when $a$ is sufficiently large, receiver 1 can decode the entire transmission from user 2 when the two users employ optimal signaling as in the single user case. However, when $a\geq 1$ but is not large enough to satisfy (\ref{eqn:condition1_1}) of Proposition \ref{thm:thm1}, the pair of interference-free rates may no longer be supportable by the channel. In such a case, at least one user must refrain from using its optimal signaling so that the resulting pair of rates can be supported. Motivated by this observation, we consider an achievable scheme which is a generalization of the scheme used in Section III to the wider case $a\geq 1$. In this scheme, user 2 sends common message and receiver 1 jointly decodes its own codeword and user 2's codeword. As in previous schemes, transmitter $i$ still employs constant power; however, user $i$ ($i=1,2$) is relaxed to reduce its transmission rate by choosing a burstiness parameter $1-\theta_j^*\leq\theta_i\leq 1$ ($j\neq i$) other than $\theta_i^*$. As in Section III, transmission is scheduled such that the duration of overlap is equal to $ \theta_1+\theta_2-1$. When $\theta_1$ and $\theta_2$ are given, in order for receiver 1 to succeed in jointly decoding both users' codewords, user 1 and user 2's rates $R_1$ and $R_2$ should be chosen to satisfy
\begin{align}
R_1 + R_2 \leq &(1-\theta_2)C(\nu_1)+(1-\theta_1)C(a\nu_2)\nonumber\\
 &+ (\theta_1+\theta_2-1)C(\nu_1+a\nu_2)) \label{eqn:macsr}
\end{align}
The right side of (\ref{eqn:macsr}) represents the maximum possible mutual information flowing from both transmitters to receiver 1 over the entire transmission period when Gaussian codebooks are used. The maximum achievable sum rate by Scheme IV can be found by solving
\begin{align}
 \max_{\theta_1,\theta_2} &\min(\theta_1 C(\nu_1) + \theta_2 C(\nu_2), (1-\theta_2)C(\nu_1)\nonumber\\
 & +(1-\theta_1)C(a\nu_2) + (\theta_1+\theta_2-1)C(\nu_1+a\nu_2))\label{eqn:srate4}
\end{align}
where $\nu_1 = P_1/\theta_1 -\epsilon_1$ and $\nu_2 = P_2/\theta_2 -\epsilon_2$,
subject to the constraints $1-\theta_2^* \leq \theta_1 \leq 1$, $1-\theta_1^* \leq \theta_2 \leq 1$ and $\theta_1 + \theta_2 \geq 1$.

Note that in Scheme IV, user $2$'s message is encoded into a codeword that spans $n\theta_2$ time slots and receiver 1 performs joint decoding over the entire $n$ time slots. By contrast, in Scheme III, receiver 1 performs joint decoding only during the overlap. In practice, joint decoding usually has a much higher decoding complexity than that of single-user decoding. Thus, Scheme IV should result in a higher overall decoding complexity compared with Scheme III. Besides, in Scheme IV, receiver 1 is required to be on when only transmitter $2$ is on.

\subsection{Scheme V: Partial overlapping time slots, constant power for user 2 and variable power for user 1 ($a<1$ only)}
In Scheme III, both users employ constant signal power levels for transmission when they are in the ``on'' state. It is indeed optimal for user 2 to do so to maximize its own rate; however, it is not clear whether this is still true in terms of maximizing the sum rate. Nevertheless, for simplification, we still assume user 2 employs constant power but allow user 1 to do power control accordingly. We study this scheme only for $a<1$. If user 2 transmits over $\theta_2$ ($1-\theta_1^* \leq \theta_2\leq 1$) fraction of time slots, then its signal power is $\nu_2 = P_2/\theta_2-\epsilon_2$. In this case, user 1 sees two parallel Gaussian channels with one having noise variance $N_{1,1} = 1$ and $n(1-\theta_2)$ channel uses and the other having effective noise variance $N_{1,2} = 1+\nu_2$ and $n\theta_2$ channel uses. Here we have used the fact that when users overlap, since $a<1$, to maximize sum-rate user 1 treats user 2's interference as noise. Within the channel with noise level $N_{1,k}$ ($k=1,2$), it is optimal to send Gaussian signals over $\theta_{1,k}$ fraction of the channel with the same power level $\nu_{1,k}$ \cite{processingcost:Massaad_Medard_Zheng04}, where $\theta_{1,k}$'s and $\nu_{1,k}$'s can be determined by a generalized water pouring solution called ``glue pouring'' in \cite{processingcost:Massaad_Medard_Zheng_TWC08}. In essence, when user 1 uses glue pouring for power control, it should start using the channel with noise level $N_{1,2}$ only if the signal-to-noise ratio's (SNR) of the channel with noise variance $N_{1,1}$ have already reached a threshold. The readers are referred to \cite{processingcost:Massaad_Medard_Zheng_TWC08} for more details on glue pouring. It is not difficult to prove that in our case $\theta_{1,1} = 1$; i.e., user 1 would make full use of all the interference-free channel uses. Depending on the power constraints and processing costs, user 1 may or may not use the channel with noise level $N_{1,2}$.

The maximum achievable sum rate by Scheme V for $0\leq a<1$ can be found by solving
\begin{align}
 \max_{1-\theta_1^* \leq \theta_2\leq 1}\theta_2 C(\nu_2)+ (1-\theta_2)C(\nu_{1,1}) + \theta_{1,2}\theta_2C(\frac{\nu_{1,2}}{1+\nu_2})
\end{align}
where $\theta_{1,2}$, $\nu_{1,1}$ and $\nu_{1,2}$ can be obtained through glue pouring.

\subsection{Numerical Results}

\begin{figure}
\centering
\includegraphics[width = 3in]{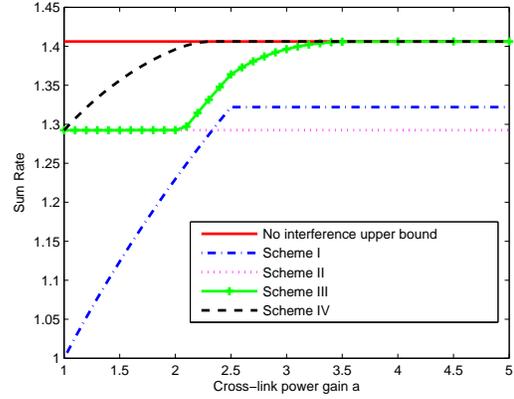}
\caption{Comparison of sum rates achieved by various schemes when $a\geq 1$}
\label{fig:fig2a}
\end{figure}

Fig. \ref{fig:fig2a} plots the sum rate achieved by the proposed schemes as a function of cross-link power gain $a$ when $P_1 = P_2 = 3.5$ and $\epsilon_1 = \epsilon_2 = 2$. Given these parameters, we have $\theta_1^* = \theta_2^* = 0.76$ and $\nu_1^* = \nu_2^* = 2.59$. Obviously, $\theta_1^*+\theta_2^*>1$. For comparison, the sum of both users' maximum interference-free rates is provided as an upper bound. Fig. \ref{fig:fig2a} shows that when $a\geq 1$ Scheme IV which employs joint decoding at receiver 1 is superior to the others and can achieve the upper bound when $a$ is greater than 2.3. Note that $a\geq 2.3$ corresponds to the very strong interference regime in Proposition \ref{thm:thm1} of Section III where interference incurs no rate loss. In contrast, in the case of no processing overhead, this regime is given by $a\geq 1+P_1 = 4.5$. Hence, the presence of processing overhead makes the very strong interference regime become larger. Scheme III can achieve the upper bound as well but for a smaller range of $a$'s compared with Scheme IV. This is because receiver 1 only jointly decodes the intended signal and the interfering signal during the overlap, which is strictly suboptimal for certain $a$'s.  As TDM does not depend on $a$, the sum rate of Scheme II remains constant. Depending on the value of $a$, Scheme I and Scheme II may outperform one another.

\begin{figure}
\centering
\includegraphics[width = 3in]{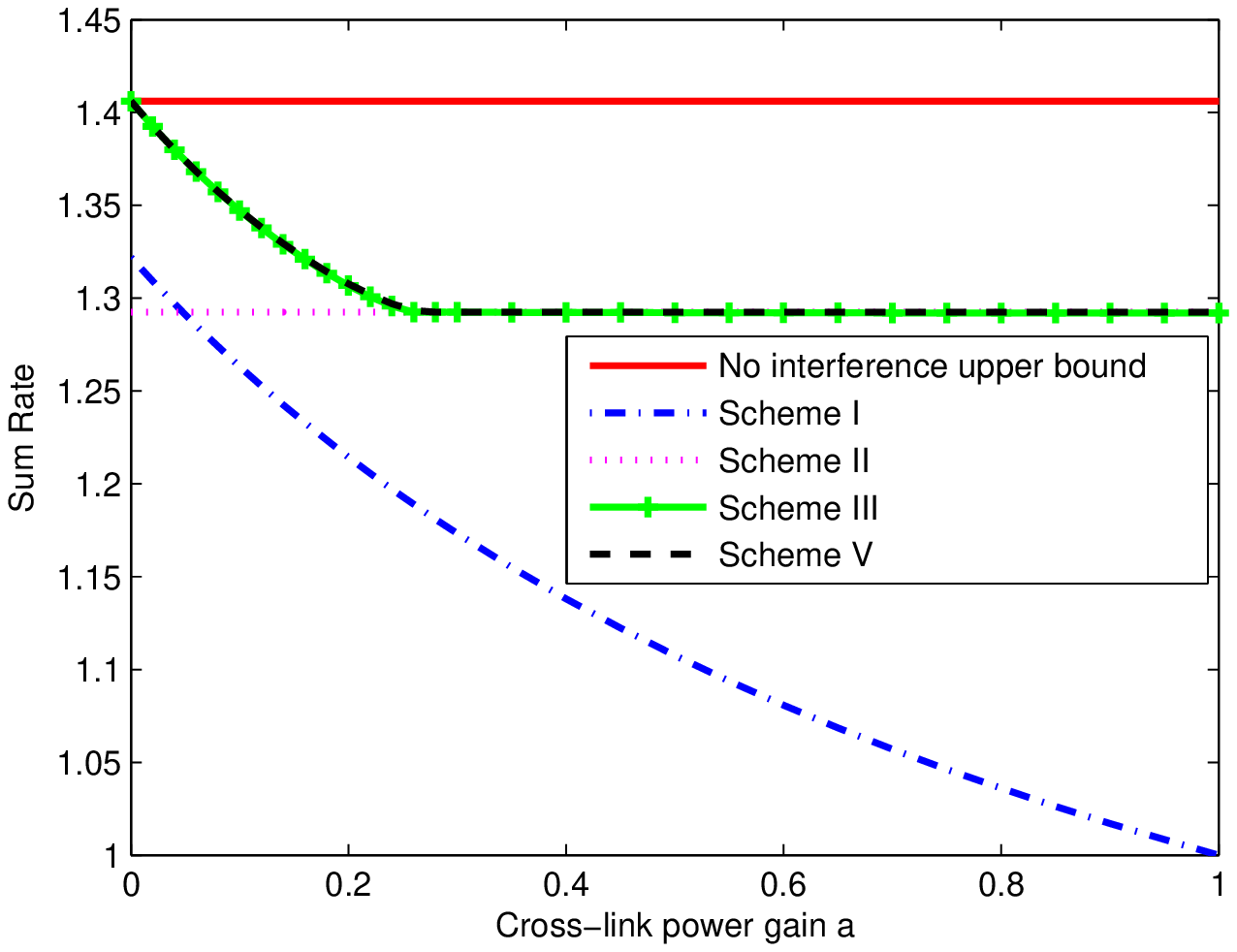}
\caption{Comparison of sum rates achieved by various schemes when $a<1$}
\label{fig:fig3}
\end{figure}

For $0 \leq a<1$, as shown in Fig. \ref{fig:fig3}, Scheme V has the best sum-rate among all four schemes and is strictly better than TDM for $a\leq 0.28$. Scheme III has extremely close performance to Scheme V and its maximum sum rate is only slightly smaller when $a$ is between 0.15 and 0.28. This seems to suggest that the flexibility we allowed in choosing burstiness parameters $\theta_1$ and $\theta_2$ in Scheme III almost diminishes the effect of no power control at transmitter 1. For moderate $a$'s in the range $0.28 <a <1$, it can be observed that Scheme III and Scheme V do not perform better than TDM in terms of sum rate. It is worth mentioning that for the MAC with processing overhead, TDM was shown in \cite{processingcost:Massaad_Medard_Zheng_Allerton} to be the best scheme in maximizing the sum rate.

\section{Conclusions}
In this paper, we assume each transmitter of a two-user Gaussian Z-IC has a constant processing energy cost per channel use and study the impact of this overhead on the capacity. With processing overhead, it is no longer optimal for each transmitter to transmit all the time as in the conventional no-overhead case.  For the Gaussian Z-IC, conditions are identified for the very strong interference regime and shown to correspond to a larger range of channel parameters than its counterpart in the no-overhead case. Assuming fixed on-off states of transmitters, several simplified joint transmission schemes are proposed to maximize the sum rate. Numerical results show that with a simple scheme one can either achieve or get very close to the upper bound for very small or large $a$'s. Future works include extension of the analysis to the two-sided Gaussian IC and derivation of bounds on capacity when on-off states are used for signaling. It would also be of practical interest to consider more accurate models for transmitter processing energy cost like the one in \cite{processingcost:Cui_Goldsmith_Bahai05}.

\bibliographystyle{IEEEtran}
\bibliography{IEEEabrv,processingcost}
\end{document}